\documentclass[11pt,a4paper]{article}
\usepackage{jheppub}

\usepackage{atlasphysics}
\usepackage{wasysym}
\usepackage{color}
\usepackage{lineno}
\usepackage{hyperref}

\def\nttb{\ensuremath{t\overline{t}}}
\newcommand{\mgg}{\ensuremath{m_{\gamma\gamma}}}
\newcommand{\tqhtitle}{{
Search for top quark decays $t\rightarrow qH$ with $H\to\gamma\gamma$ using the ATLAS detector
}}

\newcommand\tqhabstract{{
A search is performed for flavour-changing neutral currents in the decay of a top quark to an up-type ($c, u$) quark and a Higgs boson, where the Higgs boson decays to two photons. The proton-proton collision data set used corresponds to 4.7 fb$^{-1}$ at $\sqrt{s}$ = 7 \tev\ 
and 20.3 fb$^{-1}$ at $\sqrt{s}$ = 8 \tev\ collected by the ATLAS 
experiment at the LHC. Top quark pair events are searched for in which one top quark decays to $qH$  and the other decays to $bW$. Both 
the hadronic and the leptonic decay modes of the $W$ boson are used. No significant signal is observed and 
an upper limit is set on the $t\rightarrow qH$ branching ratio of 0.79\% at the 
95\% confidence level. The corresponding limit on the $tqH$ coupling combination $\sqrt{\lambda_{tcH}^{2} + \lambda_{tuH}^{2}}$
is 0.17.
}}

\usepackage{preprintcover} 
\PreprintCoverPaperTitle{\tqhtitle}
\PreprintIdNumber{CERN-PH-EP-2014-036}  
\PreprintCoverAbstract{\tqhabstract}
\PreprintJournalName{JHEP}

\begin{document}

\title{\tqhtitle}
\author{The ATLAS Collaboration}

\abstract{\tqhabstract}

\maketitle

\keywords{Top physics, Higgs boson, Flavor changing neutral interaction}

\section{Introduction}
\label{sec:introduction}
 The observation by the ATLAS~\cite{ObservationATLAS} and the CMS~\cite{ObservationCMS} Collaborations of a new
 boson with a mass around 125 \gev, compatible with the long-sought Higgs boson~\cite{higgs1,higgs2,higgs3,higgs4},
 opens up the possibility of searching for the decay of a top quark to a Higgs boson plus an up-type ($c, u$) quark. Such a decay would
 proceed via a flavour-changing neutral current (FCNC). According to the Standard Model (SM), FCNC processes are forbidden at tree level and very much suppressed at higher orders due to the 
 Glashow-Iliopoulos-Maiani (GIM) mechanism~\cite{Glashow:1970gm}. For instance, the expectation for the 
$t \to cH$ branching ratio is $\sim 3 \cdot 10^{-15}$ (see Ref.~\cite{AguilarSaavedra:2004wm} and references therein).
Observations of FCNC decays of the top quark would therefore provide
 a clear signal of new physics. 
\par Previous searches for FCNC were conducted in particular for the $t\rightarrow c(u)Z$ decay mode by  the LEP and HERA experiments~\cite{alephFCNC, delphiFCNC, l3FCNC, opalFCNC, H1:2004, Zeus:2012} (via the crossed processes),  CDF~\cite{Aaltonen:2008aaa} ,
 ATLAS~\cite{ATLAS2012:tcZ} and CMS~\cite{CMS2013:PAS}.
The current best limit for the branching ratio is 0.05\% at the 95\% confidence level~\cite{CMS2013:PAS}.

\par In models beyond the SM, the GIM suppression can be relaxed, and loop 
diagrams mediated by new bosons may contribute, yielding effective couplings $\lambda_{tqH}$ orders 
of magnitude larger than those of the SM. Examples of such extensions are the quark-singlet 
model (QS)~\cite{AguilarSaavedra:2002ns, delAguila:1998tp, AguilarSaavedra:2002kr}, two-Higgs-doublet models (2HDM) of type I, with explicit flavour conservation, and of type II, such as the minimal supersymmetric standard model (MSSM)~\cite{Li:1993mg, deDivitiis:1997sh, Lopez:1997xv, Guasch:1999jp, Delepine:2004hr, Liu:2004qw, Cao:2007dk}. In 2HDM without explicit flavour conservation (type III)~\cite{Cheng:1987rs, Grzadkowski:1990sm, Hou:1991un, Luke:1993cy, Atwood:1996vj, Bejar:2000ub,Baum:2008, Chen:2013,Atwood:2013}, the $tc(u)H$ couplings are present at tree level.
For a review of the different models see Ref.~\cite{AguilarSaavedra:2004wm}.

\par Among the published extensions of the SM, the largest branching ratio ($\sim 0.15\%$) is specific to the $t\rightarrow cH$ decay. It appears in 2HDM of type III
and corresponds to a non-flavour-diagonal Yukawa coupling which scales with top-quark and light-quark masses, $m_t$ and $m_q$, as 
$\lambda_{tqH} = \sqrt{2 m_qm_t}/v$, as proposed in Ref.~\cite{Cheng:1987rs}, where $v/\sqrt{2} = 174 \gev$ is 
the Higgs field vacuum expectation value. 
In the other models discussed in Ref.~\cite{AguilarSaavedra:2004wm}, the largest branching ratios for the $t \rightarrow qH$ decays are of the order of a few 10$^{-5}$.

\par In this paper a search for $t\rightarrow qH$ decays in \nttb\ production is undertaken
using the $H\rightarrow\gamma\gamma$  decay channel. The analysis does not distinguish between the $t\rightarrow cH$ and $t\rightarrow uH$ final states
which have similar acceptances. As theory favours $t\rightarrow cH$, this mode is used as reference throughout this work, unless otherwise stated.
Despite its small branching ratio ($\sim 0.23 \%$ for a Higgs boson mass around 125 \gev), the $H \rightarrow\gamma\gamma$ channel 
was chosen because of its demonstrated high importance for inclusive Higgs boson studies, with a rather large number of events
and a clean signature~\cite{ObservationATLAS,Aad:2013wqa}. 
The remaining top quark in the event is searched for in two final states: a bottom quark and a hadronically decaying $W$ boson, giving rise  
to events with four jets, or a leptonically decaying $W$ boson, giving two jets, a lepton and missing transverse energy.
\par The branching ratio ${\cal B}$ of the $t \to qH$ process is estimated as the ratio of its partial width to 
the $t\rightarrow bW$ width, assumed to be dominant,
\begin{eqnarray}
{\cal B} = (\lambda^2_{tcH}+\lambda^2_{tuH})/(g^2 \cdot |V_{tb}|^2 \cdot \chi^2),
\label{eq1}
\end{eqnarray}
where $|V_{tb}|$ is taken equal to 1, $\chi$ is a kinematic factor~\footnote{$\chi^2 = (1 - 3x^4 + 2 x^6)(1 - y^2)^{-2}x^{-2}/2$, 
where $x = m_W/m_t$, $y = m_H/m_t$, $m_W$ and $m_H$ are the $W$-boson and Higgs boson masses
and the masses of the other quarks are neglected.}
and $g = 2m_W/v$ is the weak coupling constant. Using PDG averages~\cite{PDG:2012} and 
applying NLO corrections to both the $t\rightarrow qH$ partial width~\cite{Zhang:2013xya} and the top quark 
total decay width~\cite{TopWidth:1991} leads to $\chi g = 1.92 \pm 0.02$, which is used in the extraction 
of the coupling. 

\section{Detector and data set}
\label{sec:sample}

\par The ATLAS detector~\cite{ATLAS_detector} consists of an inner tracking detector (ID)
surrounded by a superconducting solenoid providing a 2\,T magnetic field,
electromagnetic and hadronic calorimeters, and a muon spectrometer. The ID provides tracking in the pseudorapidity\footnote{ATLAS uses a
right-handed coordinate system with its origin at the nominal interaction point
(IP) in the centre of the detector and the $z$-axis along the beam line. Observables labelled as transverse are projected onto the $xy$ plane. The
$x$-axis points from the IP to the centre of the LHC ring, and the $y$-axis
points upwards. Cylindrical coordinates $(r, \phi)$ are used in the transverse
plane, $\phi$ being the azimuthal angle around the beam line. The pseudorapidity
is defined in terms of the polar angle $\theta$ as $\eta = -\ln\tan\frac{\theta}{2}$. The $\Delta R$ distance is defined as
$\Delta R = \sqrt{(\Delta \eta)^2 + (\Delta \phi)^2}$.}
 region $|\eta| < 2.5$ and consists of silicon pixel- and microstrip-detectors
 inside a transition radiation tracker. 
The electromagnetic calorimeter, a lead/liquid-argon sampling device, is divided into
 one barrel ($|\eta|<1.475$) and two end-cap ($1.375<|\eta|<3.2$) sections.
 Longitudinally, it is divided into three layers. The first layer, referred to as
 the strip layer, has a fine segmentation in the regions $|\eta|<1.4$ and $1.5<|\eta|<2.4$ to facilitate the separation
 of photons from neutral hadrons and to allow shower
 directions to be measured, while most of the energy is deposited in the second layer. In the range of $\lvert\eta\rvert < 1.8$ a
 presampler layer inside the cryostat allows for the correction of energy losses
 upstream of the calorimeter. The barrel ($|\eta|<1.7$) hadronic calorimeter consists
 of steel and scintillating tiles, while the end-cap sections ($1.5<|\eta|<3.2$)
 are composed of copper and liquid argon. The forward calorimeter
 ($3.1<|\eta|<4.9$) uses copper and tungsten as absorber with liquid argon as active material.
The muon spectrometer consists of precision ($|\eta|<2.7$) and trigger ($|\eta|<2.4$) chambers embedded in a toroidal magnet system which surrounds the hadronic calorimeter.

\par This analysis uses the full proton-proton data set recorded by ATLAS in 2011 and 2012. After application of data-quality requirements, the integrated luminosity amounts to 4.7 fb$^{-1}$ at $\sqrt{s}=7$ \tev, with a relative uncertainty of 1.8\%~\cite{lumi:2011}, and 20.3 fb$^{-1}$at  $\sqrt{s}=8$ \tev, with a relative uncertainty of 2.8\%\footnote{The luminosity of the 2012 data set is derived, following the same methodology as that detailed in Ref. ~\cite{lumi:2011}, from a preliminary calibration of the luminosity scale derived from beam-separation scans performed in November 2012.}.
The data were recorded with instantaneous luminosities varying between $1\times
10^{32}\,\text{cm}^{-2}\text{s}^{-1}$ and $7.8\times 10^{33}\,\text{cm}^{-2}\text{s}^{-1}$.
The mean number of interactions per bunch crossing was 9.1 in 2011 and 20.4 in 2012. 
The inelastic collisions that occur in addition to the hard interaction produce mainly low transverse momentum particles that form the so-called ``pile-up'' background.

\par The data considered here were selected using a diphoton trigger in which two clusters formed from energy depositions in the electromagnetic calorimeter
are required. A transverse energy ($E_\mathrm{T}$) threshold of 20 \gev\ was required at 7 \tev, while at 8 \tev\ the thresholds were increased to 35 \gev\ and 25 \gev\
on the leading (sorted in $E_\mathrm{T}$)
and sub-leading clusters. In addition, loose criteria were applied on the shape of the clusters requiring that they match the expectations for electromagnetic showers initiated by photons. 
For events satisfying the off-line selection, the efficiency of the diphoton trigger is measured to be $(98.9\pm 0.2)\%$ for $\sqrt{s}=7$ \tev\ data and $(99.6\pm 0.5)\%$ at 8 \tev.
\section{Experimental techniques}
\subsection{Photon reconstruction and identification}
\label{sec:photonID}
The photon reconstruction is seeded from clusters of energy deposits in the
electromagnetic calorimeter. 
Clusters without any matching track in the ID are classified as
unconverted photon candidates. Clusters with a matching conversion reconstructed from one
or two tracks are classified as converted photon candidates~\cite{Aad:2013zba}.
The efficiency of the photon reconstruction is about 96.5\% averaged over
the $E_\mathrm{T}$ and $\eta$ spectra expected for photons from a $m_H = 125$ \gev\  Higgs boson decay.

The identification of photons (PID) is based on the shape of their showers in the electromagnetic calorimeter. 
An initial loose selection, also used at trigger level, is based on shower shapes in the second layer of the 
electromagnetic calorimeter and on the energy deposition in the hadronic calorimeter.
The tight identification adds information from the
finely segmented strip layer.
 The PID efficiency, averaged over $\eta$, ranges between
85\% and 95\% for the $E_\mathrm{T}$ range of interest.
\label{sec:pidunc}
\par The measurement of the uncertainty on the PID efficiency is based on the comparison of the efficiency 
obtained in the simulation and the combination of three data-driven measurements~\cite{PHOTONID2011}. Taking into account possible correlations in $\eta$, $E_\mathrm{T}$ and conversion status, the resulting uncertainty on
 the diphoton inclusive signal yield is estimated to be 8.4\% at 7 \tev\ and, owing to the larger sample and several analysis improvements, 2.4\% at 8 \tev~\cite{Aad:2013wqa}.
For the hadronic channel analysis, where four or more jets are required, the systematic uncertainty  is 9.3\% at 7 \tev\ and 4.6\% at 8 \tev. 
\par To further suppress jets faking photons,  calorimetric and track isolation requirements are applied.
The isolation-$E_\mathrm{T}$ 

is estimated by summing the  $E_\mathrm{T}$ of
positive-energy topological clusters\footnote{Topological clusters are three-dimensional
clusters of variable size, built by associating calorimeter cells on the basis
of the signal-to-noise ratio~\cite{TopoClusters}.} reconstructed in the
electromagnetic and hadronic
calorimeters in a cone of $\Delta R = 0.4$ around the photon candidate, where
the region of size $0.125\times 0.175$ in $\eta\times\phi$ around the photon
barycentre is excluded. The isolation-$E_\mathrm{T}$ is corrected for leakage of the
photon energy outside of the excluded region and for pile-up~\cite{Aad:2010sp}, and
it is required to be below 4 \gev\ (6 \gev) for the 2011 (2012) data. For the 2012 data set, the scalar sum of the transverse momentum ($p_\mathrm{T}$) of all tracks consistent with the primary vertex (see below), with $p_\mathrm{T}$ above 1~\gev\ and 
in a $\Delta R = 0.2$ cone around the photon direction is required to be below 2.6 \gev. Comparing data and
simulation using electrons from $Z\to e^+ e^-$ candidate events, and photons from
$Z\to \ell^+ \ell^-\gamma$ events, where $\ell = e$ or $\mu$, a good agreement between efficiencies is found and the remaining small difference is accounted for as a
systematic uncertainty of 1$\%$ on the diphoton signal yield for inclusive production. For events with four or more jets, the efficiency of the calorimetric isolation selection was found to be slightly smaller in data than
in the simulation, resulting in a correction factor of 0.98 with a systematic uncertainty of $\pm$3$\%$.

\par The energies of the clusters are calibrated separately for unconverted and
converted photon candidates and for electrons.

The energy calibration for data is refined by
applying $\eta$-dependent correction factors, which are about
$\pm 1\%$, determined from $Z\to e^+ e^-$ events. The simulation is corrected to
reflect the energy resolution observed using $Z\to e^+ e^-$ events in data,
which requires an energy smearing of about $1\%$ in the 
barrel region and between $1.5\%$ and $2.5\%$ in the end-cap region of the calorimeter.

\subsection{Other physics objects}

\label{sec:PV}
\par  
 The kinematic properties of all objects are determined with respect to a primary vertex selected~\cite{Aad:2013wqa} by combining:

- an estimate of its $z$ position obtained from the intersection of the beam line with the direction of the photons, as determined 
by the measurement using the longitudinal segmentation of the calorimeter, and the conversion point or hits 
in the ID when available;

- the scalar sum of the transverse momenta and the sum of the squared transverse momenta 
of the tracks associated with each reconstructed vertex;

- and, at 8 \tev, the difference in azimuth between the direction of the vector sum of the tracks momenta and 
the di-photon system.

\label{sec:objects}
In addition to photons, the analysis requires also jets, electrons or muons and missing transverse 
energy, $\met$. The main inputs to identify and measure these objects are summarised below.

\begin {itemize}
\item 
\label{sec:jet}
Jets are reconstructed from topological clusters in the calorimeters~\cite{TopoClusters,JESPaper2010}, using the anti-$k_t$ algorithm~\cite{JetAlgo} with a radius parameter $R=0.4$. They must
have $|\eta|<4.5$ and $p_\mathrm{T}>25$ \gev. For the 8 \tev\ data set, this threshold is increased to 30 \gev\ for jets with $|\eta|>2.4$. The dependence of the jet response on the number of reconstructed primary vertices and on the expected average number of interactions is removed, at $\sqrt{s}=8$ \tev\ where the pile-up is largest, 
by applying an event-by-event subtraction procedure based on the jet area method~\cite{jetarea}.
In order to suppress jets produced in additional pile-up interactions, each jet is also
required to have a sufficiently high jet vertex fraction (JVF) defined as the scalar sum of $p_\mathrm{T}$ of the tracks consistent with the primary vertex that fall into the jet area over the sum of track $p_\mathrm{T}$ 
from all primary vertices falling into the same jet area.
A JVF larger than 0.75 (0.25) for the 7 \tev\ (8 \tev) data set is required. At 8 \tev, this cut is only applied for jets with $|\eta|<2.4$ and $p_\mathrm{T}<50$ \gev. 

\item 
\label{sec:Btagging}
 
The tagging of bottom quark jets is performed using a neural network identifier~\cite{ATLAS-Btagging}, which includes information from the impact parameter of tracks and from displaced vertices from hadron decays. The threshold values are set so as to give, on average, a 70\% efficiency 
for jets containing a bottom hadron in \nttb\ events. The efficiency for charm jets is about 20\%, and it is less than 1\% for light-quark jets. Small differences between data and simulation are taken into account by a global factor determined by propagating to the simulated signal samples the differences between data and simulation measured on dedicated samples (\nttb\ in particular).

\item 
\label{sec:ele}
Electron candidates consist of clusters of energy deposited in the electromagnetic 
calorimeter that are associated with ID tracks~\cite{ElectronPerformance}. Their  
transverse energy is computed from the cluster energy and the track direction at the 
interaction point, and they are required to satisfy $|\eta|< 2.47$ and $E_\mathrm{T} > 15$ \gev.

Electron candidates have to pass a set of requirements 
on the hadronic leakage, shower shapes, track quality and track-cluster matching variables. 
Furthermore, they must be isolated: the calorimetric isolation  $E_\mathrm{T}$ in a $\Delta R = 0.4$ cone divided by the electron candidate's $E_\mathrm{T}$ is required to be 
less than 0.2, and the scalar sum of the $p_\mathrm{T}$ of tracks consistent with the primary vertex, in a cone of $\Delta R = 0.2$ around the electron 
candidate's track, divided by its $E_\mathrm{T}$, has to be less than 0.15.

\item 
\label{sec:muon}
Muon candidates are required to pass the conditions $|\eta|< 2.7$ and $p_\mathrm{T} > 10 \gev$, and they must be isolated with the same criteria as for electrons.
The muon tracks must have a transverse impact parameter $|d_0| < 1$ mm and a longitudinal impact parameter $|z_0| < 10$ mm.

\item 
\label{sec:met}
The missing transverse energy is calculated as the magnitude of the sum of the $\vec{p}_\mathrm{T}$ of all identified objects in an event. Clusters of calorimeter cells with $|\eta|<4.9$ not associated with any of the objects described above are also added.
 \end{itemize}

\par A given particle may be reconstructed as more than one object (for example both as a photon and a jet). This possible duplication, as well as any real overlap in a narrow $\Delta R$ interval, is suppressed by an overlap removal procedure (within $\Delta R = 0.2$ between photons and electrons, 0.4 between photons and jets, 0.4 between muons and either jets or photons) in which the highest priority is given to photons, followed by electrons, jets and finally muons. 

\subsection{Signal and background simulation}
\label{sec:MC}
The simulations of the signal and of the SM Higgs boson production (resonant background) are used to estimate 
the corresponding acceptances. The relevant non-resonant backgrounds  are due to diphoton 
production and \nttb\ and $W$ production. These backgrounds are simulated in order to constrain the shape 
of the fitted non-resonant background in control regions of the data.

 \par Signal events corresponding to \nttb\ production, with one top quark decaying
into a charm quark and a Higgs boson (which is constrained to decay into two photons) 
were generated using PROTOS 2.2~\cite{PROTOS:2009}, with PYTHIA6~\cite{PYTHIA6} for
 parton shower (PS), multiple interactions (MI) and hadronisation, with a set of parameters as defined by the Perugia2011C tune~\cite{Skands:2010ak}.
A top quark mass of 172.5 \gev\ and a Higgs boson mass of 125 \gev\ were chosen. 

Four samples of 30,000 events were generated: two samples correspond to $\sqrt{s}=8$ \tev\ and the other two
to  $\sqrt{s}=7$ \tev. At each energy there is one sample for which
the second top quark decays only hadronically
and one sample where the $W$ boson from the second top quark decays leptonically, including tau leptons which are decayed using TAUOLA~\cite{TAUOLA:1990}. The hadronic and the leptonic samples are added with weights corresponding to the respective decay fractions of the $W$ boson. At 8 \tev, two additional samples were generated where one top quark decays to an up quark (instead of a charm quark) and a Higgs boson, which allow for the determination of the ratio of the acceptances of the $t\rightarrow uH$ and $t\rightarrow cH$ decay modes. 

\par The contributions of known SM sources of Higgs boson
production are estimated simulating Higgs boson production by gluon fusion ($ggF$), by vector boson fusion (VBF),
Higgs-strahlung associated production ($WH$ and $ZH$), and associated production of Higgs boson 
and a \nttb\ pair (\nttb$H$). The first two were produced using
POWHEG~\cite{POWHEG:2009, POWHEG:2010} interfaced to PYTHIA8~\cite{PYTHIA8} for the Higgs boson decay, PS, MI and hadronisation, and the last three by PYTHIA8. 
\par Non-resonant production of two-photon final states  with several additional jets  dominates the background in the hadronic selection. This was simulated using SHERPA~\cite{SHERPA:2009} with up to three additional partons in the
primary hard interaction (this sample, of about $10^{7}$ events, is called S$_{\gamma\gamma j}$ in the following). 
The same final state where one of the photons is a fake photon candidate resulting from jet misidentification also contributes to the hadronic background 
(see Refs.~\cite{Aad:2013wqa, Aad:2011ww}).  The level of this additional contribution, for final states with several jets, is estimated by data-driven methods to be about 18\% of the background with two real photons. Comparative studies using dedicated simulations for the hadronic background with fake photon candidates show that, within the limited statistical precision  of these simulations due to the high rejection power of the photon identification, the distributions relevant for the hadronic analysis (see Sec.~\ref{sec:FullHadronic}) are compatible with those from the  S$_{\gamma\gamma j}$  sample. The latter is thus used in the following to represent both backgrounds.
\par Finally samples of \nttb\ ($\sim 1.5\cdot 10^{7}$ events) and $W(\gamma)$ ($\sim 2.3\cdot 10^{7}$ events) production simulated with MC@NLO~\cite{MC@NLO} and ALPGEN~\cite{ALPGEN}, respectively, interfaced to HERWIG~\cite{HERWIG} and JIMMY~\cite{JIMMY}, were used to estimate the contribution of these processes. 

The  $W(\gamma)$ sample is a $W$ sample in which the production of one photon at the matrix element level is imposed, and a lepton filter is applied.
\par In all samples but the S$_{\gamma\gamma j}$ one, PHOTOS~\cite{Golonka:2005pn} is employed to describe additional photon radiation from charged leptons.
\par All samples were processed through a full simulation of the ATLAS detector~\cite{ATLAS:SIM} based on the GEANT4 package~\cite{Geant4:2003}. A modelling of the event pile-up from the same and nearby bunch crossings, tuned to the data, is also included. The simulations are corrected using weights to reflect the number 
of interactions per bunch crossing and the spread of the $z$ position of the primary vertex observed 
in data. Differences in efficiencies between data and simulation for object reconstruction and identification 
are corrected in the same way.

\section{Analysis strategy and candidate event selection}
\label{sec:eventsel}
 Events are first required to fulfill the criteria used for the Higgs boson analysis in the
$\gamma\gamma$ channel~\cite{Aad:2013wqa}, namely to
 contain at least two reconstructed photon candidates in the fiducial region of the calorimeter 
$|\eta|<2.37$, but excluding the transition region between the barrel and endcap calorimeters, $1.37<|\eta|<1.56$. The leading (subleading) photon candidate is required to 
have $E_\mathrm{T}>40 \gev$ ($30\gev$). Tight identification and isolation criteria, as described above, are applied to both photon candidates. 
\par Additional requirements are applied in order to select events 
 compatible with a \nttb\ intermediate state.

Finally the diphoton mass distribution of the
selected events is analysed using a sideband technique in order to estimate the background in the  signal region.
\par  In the $\sqrt{s} = 8$ \tev\ data set, candidates which contain one and only one lepton are treated in the leptonic analysis
while events having no leptons are treated in the hadronic analysis, and events with two or more leptons are rejected. 
At 7 \tev\ no analysis in the leptonic channel was performed. Due to the smaller data sample and the lower sensitivity of the leptonic channel compared to the hadronic one, this has no significant impact on the precision of the final results.
\subsection{Selection of hadronically decaying top quarks}
\label{sec:FullHadronic}

Events are required to have at least four jets among which at least one is $b$-tagged. In case
of more than four jets, the four leading ones (ordered in decreasing $p_\mathrm{T}$) are considered.
However, the jet ranked 4$^{th}$ in $p_\mathrm{T}$ is replaced by the 5$^{th}$ one if the former is 
not $b$-tagged and the latter is. This procedure is extended to the 6$^{th}$ jet if the 5$^{th}$ is 
not $b$-tagged either. The signal sample shows that such a jet replacement happens for about 6\% of the 
events, and that the acceptance is increased by about the same amount. 
\par After the selection of four jets, one top-quark candidate, {\it Top1}, is constructed from the two photons and one jet, and another top-quark candidate, {\it Top2}, is formed from the three remaining jets. At least one of the four possible pairs must have masses $m_1 \equiv m_{\gamma\gamma j}$ and $m_2 \equiv m_{jjj}$ that lie within certain mass windows of size $\Delta m_1$ and $\Delta m_2$ around the top-quark mass (see below). Additional requirements, such as associating the $b$-tagged jet with {\it Top2} and imposing the invariant mass of the remaining two jets of {\it Top2} to be compatible with the $W$-boson mass were considered but not retained as they did not significantly improve the expected significance. 

\begin{figure}[h!]
\begin{center}
\hspace{-0.08cm}
\includegraphics[width=.5\textwidth]{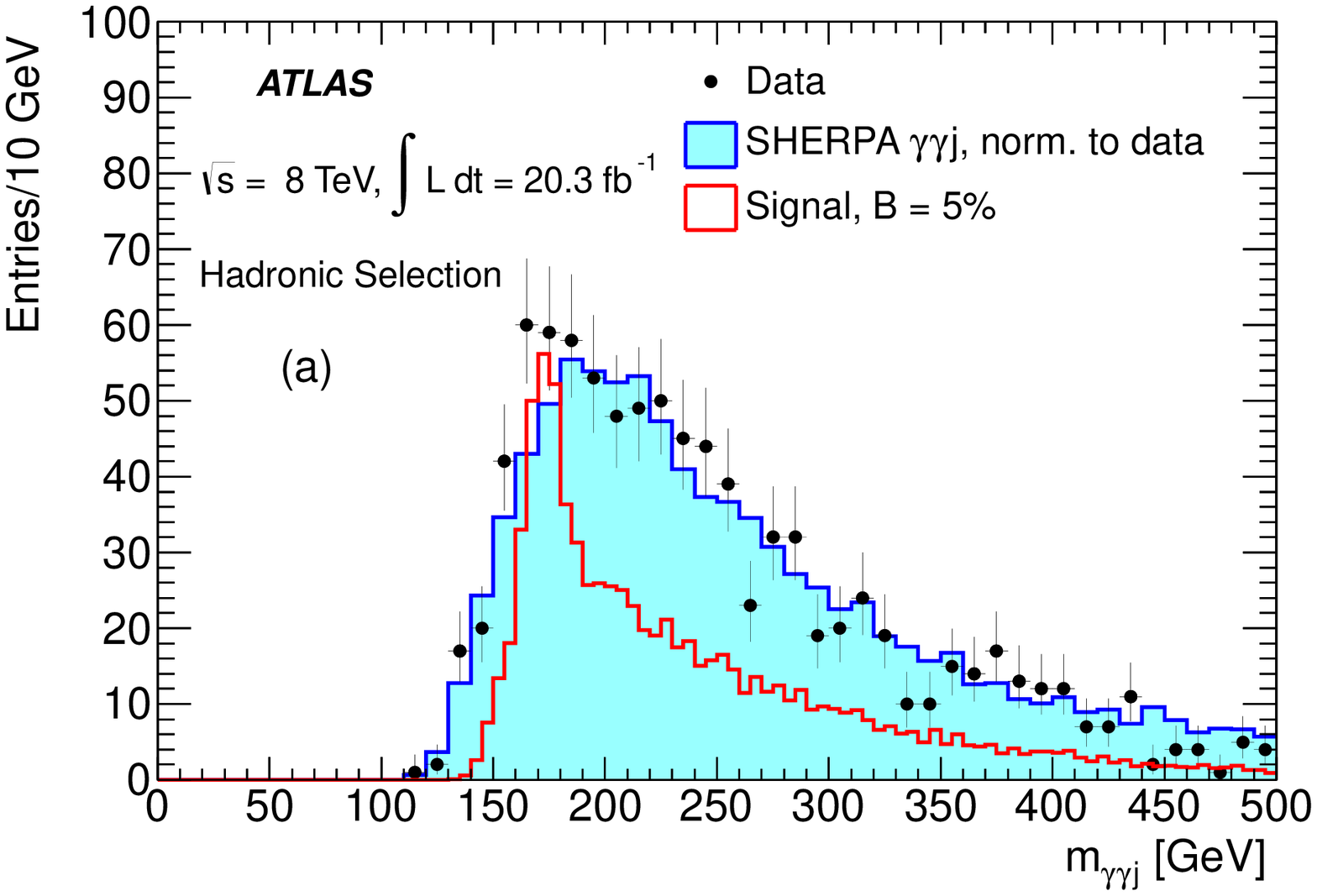}
\hspace{-0.4cm}
\includegraphics[width=.5\textwidth]{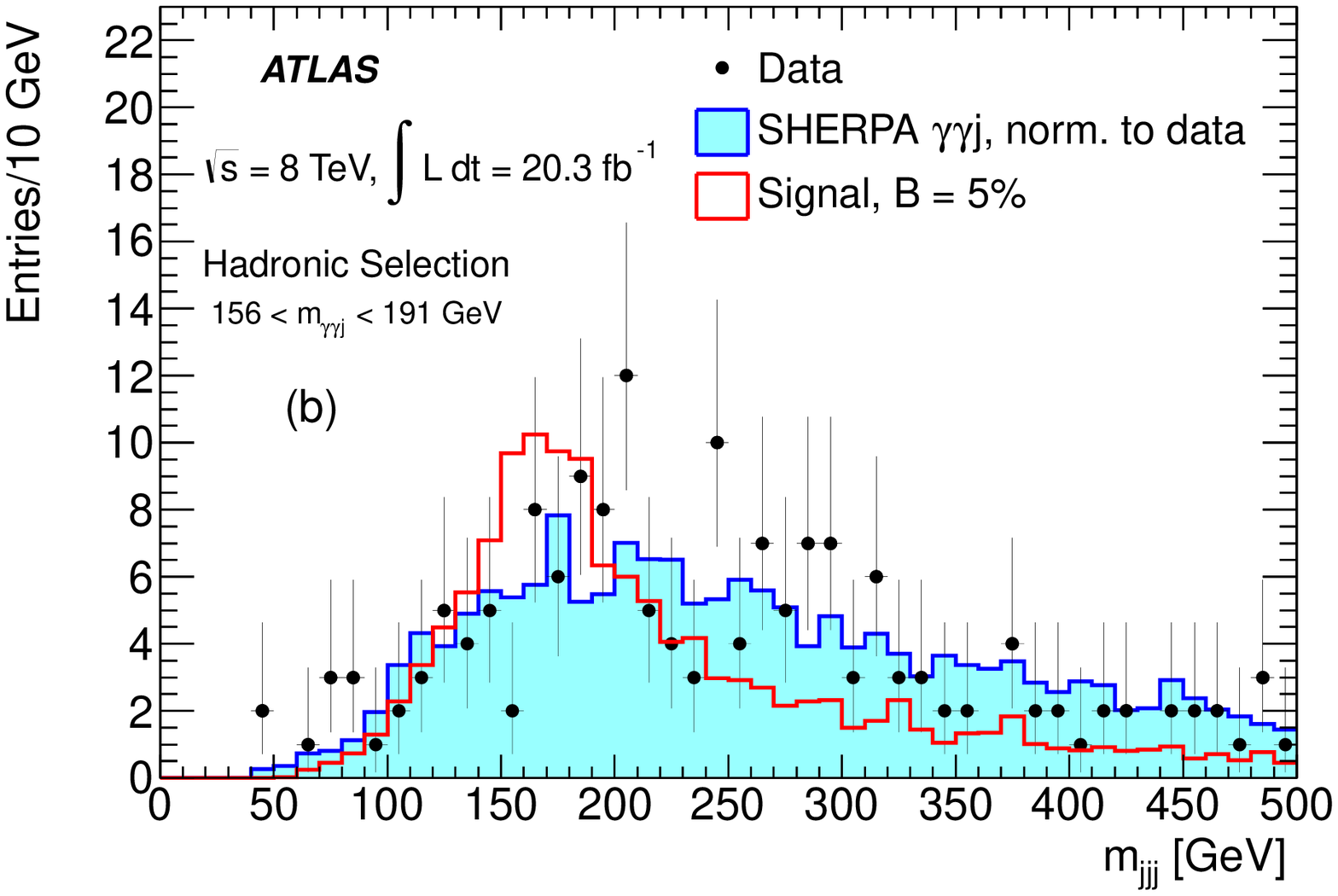}

\end{center}
\vspace{-0.5cm}
  \caption{(a) Distribution of the invariant mass $m_{\gamma\gamma j}$ ({\it Top1}),  for events selected by the hadronic analysis (see text for details) in the $\sqrt{s}=8$ \tev\ data set, with at least one $b$-tagged jet. For each event there are four  $m_{\gamma\gamma j}$ combinations, and all four are displayed. (b) Distribution of the invariant mass $m_{jjj}$ for the complementary top candidates ({\it Top2}) decaying into 3 jets; only combinations for which $m_{\gamma\gamma j}$ is between 156 and 191 \gev\  enter in the distribution.}
  \label{hadronic:fig:Mt124BT}
\end{figure}

Figure~\ref{hadronic:fig:Mt124BT}(a) shows the distribution of $m_1$ (four entries per event) for all selected 
events before mass cuts in the $\sqrt{s}=8 \tev$ data set.

In the simulated signal sample, normalised to the expectation for an arbitrary 5\% $t \rightarrow cH$ branching ratio, 
the narrow peak associated with the top quark 
is clearly visible. The combinatorial background has a shape similar
 to the distribution obtained with the S$_{\gamma\gamma j}$ sample (normalised to data in the mass region [0,500] \gev).
The background from \nttb\ and $W(\gamma)$ production is negligible.
The chosen $\Delta m_1$ interval is [156,191] \gev. 
\par The distribution of $m_2$ is shown in Fig.~\ref{hadronic:fig:Mt124BT}(b).
Only combinations for which $m_1$ fulfills the $\Delta m_1$ condition enter in this figure. The simulated signal distribution shows that 
the peak associated with {\it Top2} is broader than for {\it Top1}. The combinatorial background 
has a shape similar to the distribution obtained with the S$_{\gamma\gamma j}$ sample (normalised as for the {\it Top1} case).
The chosen $\Delta m_2$  interval is [130,210] \gev. The $\Delta m_1$ and $\Delta m_2$ 
intervals are determined on the basis of the mass resolutions observed in the simulation. 
The expected significance is stable with respect to moderate variations of the mass criteria around the chosen values.
The reconstructed mass distributions of top candidates at $\sqrt{s}=7$ \tev\ are similar to the ones shown 
at 8 \tev, and the same mass intervals are used. 

An overview of the hadronic selection at various stages of the analysis is shown in Table~\ref{tab:hadronic:stat} for both 
the $\sqrt{s}=7$ \tev\ and $\sqrt{s}=8$ \tev\ samples. 

\begin{table}[h!]
\begin{center}
\caption{Efficiency (in percent) for $t\rightarrow cH$ signal simulation and number of events for data, at  $\sqrt{s}=7$ \tev\ and  $\sqrt{s}=8$ \tev\ for the hadronic selection, at different stages of the event selection. 
The uncertainties on the efficiencies for the full selection, shown in the last row, are statistical only.}
\label{tab:hadronic:stat}
\footnotesize
\vspace*{1em}
\begin{tabular}{|l|cc|cc|}
 \hline
                &   \multicolumn{2}{c|}{$t\rightarrow cH$ ($\%$)} &  \multicolumn{2}{c|}{Data (events)}  \\ \hline \hline
                          & 7 \tev\ & 8 \tev\ & 7 \tev\ &  8 \tev\ \\ \hline \hline
 $\gamma\gamma$ selection & 34.5    & 34.2    & 23683   & 118500 \\
 N$_\mathrm{jets}\geq$ 4   & 15.2    & 15.1    &   227   &   1349 \\ 
 Mass requirements                &  5.9    &  6.1    &    36   &    210 \\
 At least 1 $b$-tag       &  4.2$\pm0.1$ & 4.0$\pm0.1$ & 7 & 43 \\ \hline
\end{tabular}
\end{center}
\end{table}  
\normalsize

\begin{figure}[h!]
\begin{center}
\includegraphics[width=.60\textwidth]{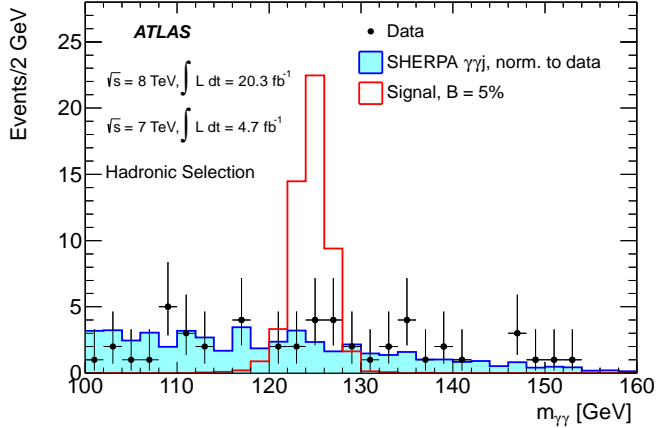}
\end{center}
\vspace{-0.5cm}
  \caption{Distribution of the invariant mass of the two photons, $m_{\gamma\gamma}$, for events passing 
the full hadronic selection (see text for details). The $S_{\gamma\gamma j}$ background sample is normalised to 
data.}
  \label{hadronic:fig:Mgaga}
\end{figure}

The $\gamma\gamma$ mass ($m_{\gamma\gamma}$) spectrum for data after the complete selection is shown in Fig.~\ref{hadronic:fig:Mgaga} together with the corresponding distribution
for the S$_{\gamma\gamma j}$ sample. The latter shows a satisfactory modeling of the background outside the expected signal mass range, as is also the case at earlier stages of the analysis with much larger statistics.  

\subsection{Selection of leptonically decaying top quarks}
\label{sec:Leptonic}

\begin{figure}[h!]
\begin{center}
\hspace{-0.08cm}
  \includegraphics[width=.49\textwidth]{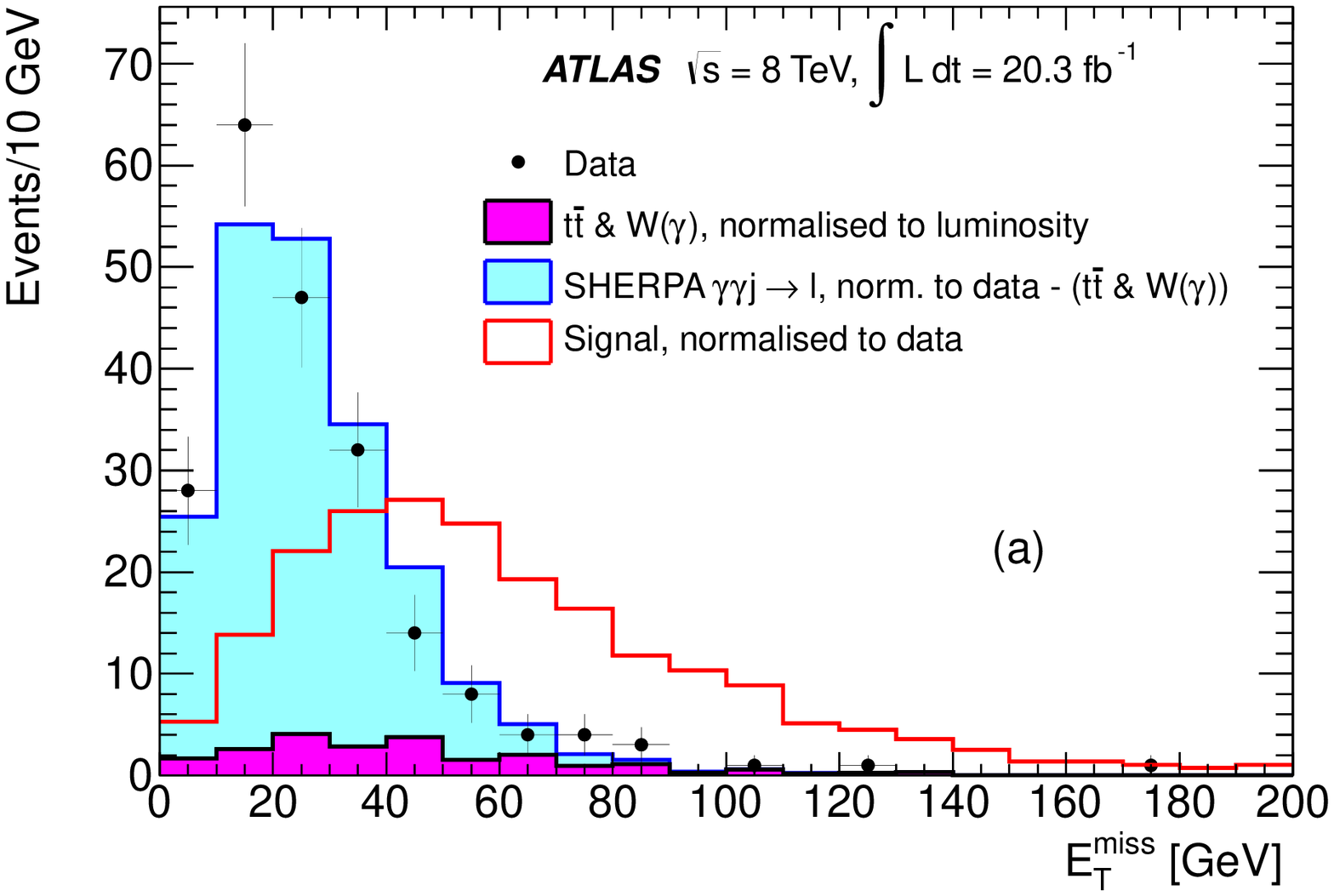}
\hspace{-0.4cm}
  \includegraphics[width=.49\textwidth]{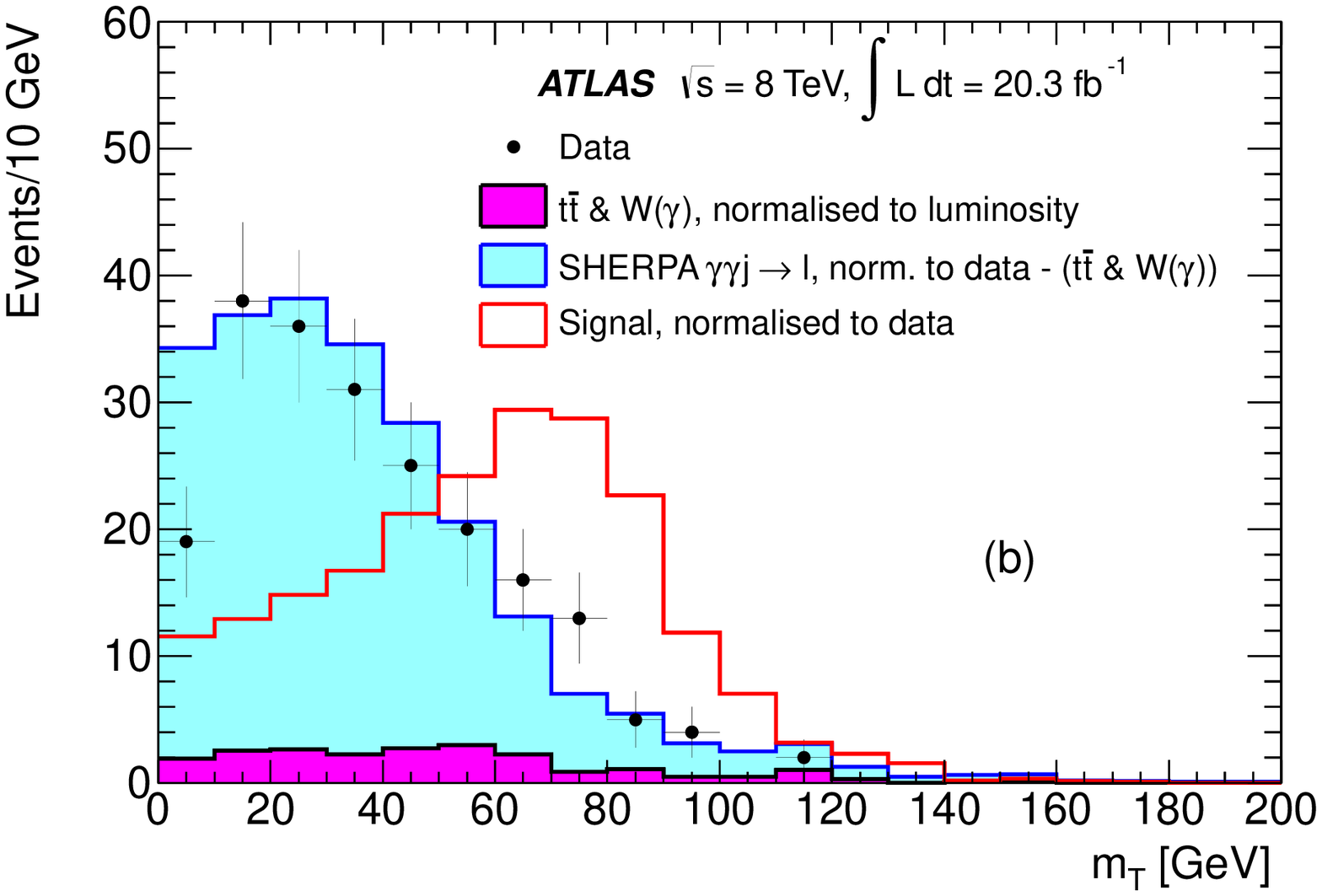}
\end{center}
\vspace{-0.5cm}
  \caption{Distributions of (a) the missing transverse energy \met\ and (b) the transverse mass $m_\mathrm{T}$ of  the $W$ candidates 
for events with two high $p_\mathrm{T}$ photons and one lepton. The  \nttb $\& W(\gamma)$ and S$_{\gamma\gamma j\rightarrow \ell}$ background samples are defined in the text.}
  \label{leptonic:fig:ETmissMT}
\end{figure}

The aim of the leptonic analysis is to identify candidate events in which 
the $W$ boson from the second top quark decays leptonically. Only electrons and muons are considered
 as identified leptons, and only events with exactly one lepton are considered.
Events with two or more jets are retained for the subsequent steps. The lepton $p_\mathrm{T}$ is used together
with \met\ to calculate the transverse mass $m_\mathrm{T}$ of the $W$ candidate, 
and $m_\mathrm{T}>$ 30 \gev\ is required.
At this stage, no event is selected in the S$_{\gamma\gamma j}$ sample, due to the high rejection power of the electron and muon 
identification requirements. In order to have a larger event sample to represent this background, one randomly chosen jet per event, 
among jets with $|\eta|<2.5$ and $p_\mathrm{T} > 15 \gev$, was replaced by a lepton with the same momentum vector. 
This sample, named S$_{\gamma\gamma j\rightarrow \ell}$, gives a good description of the data, as shown in 
Fig.~\ref{leptonic:fig:ETmissMT}(a) for \met\ and Fig.~\ref{leptonic:fig:ETmissMT}(b) for $m_\mathrm{T}$. In 
Fig.~\ref{leptonic:fig:ETmissMT} the sample referred to as \nttb $\& W(\gamma)$ originates from the \nttb\ and 
$W(\gamma)$ simulations, normalised to the luminosity of the data set, while the S$_{\gamma\gamma j\rightarrow \ell}$ sample is 
normalised to data, after subtraction of the \nttb $\& W(\gamma)$ background. 
\par The two leading jets are considered. However, as for the hadronic selection, some priority is given to $b$-tagged jets: if the jet ranked second is not $b$-tagged and if there is a $b$-tagged third jet that passes all other requirements, the second jet is replaced by the third. The replacement procedure is repeated in case there is a 4$^{th}$ $b$-tagged jet and the second and third were not $b$-tagged. The signal simulation shows that such a jet replacement happens for about 9\% of the events and that the acceptance is increased by about the same amount.

\par After the above selection of two jets, one top-quark candidate, {\it Top1}, is constructed from the two photons and one jet; its invariant mass is $m_1 \equiv m_{\gamma\gamma j}$. Another top-quark candidate, {\it Top2}, is built from the remaining selected jet, the lepton and the neutrino, with invariant mass $m_2 \equiv m_{\ell\nu j}$. The longitudinal momentum of the neutrino is estimated using a $W$-mass constraint. In the case of two real solutions~\footnote{In case no real solution exists, the constraint is applied by
replacing $m_W$ by $m_\mathrm{T}\ +$ 100 \mev, which ensures two, almost degenerate, real solutions.}, 
the one giving $m_{2}$ closer to the top-quark mass is chosen.
At least one of the two possible ({\it Top1, Top2}) pairs must have masses that lie within certain windows around the top-quark mass (see below).
Furthermore it is required that at least one of the two jets is $b$-tagged.

\begin{figure}[h!]
\begin{center}
\hspace{-0.08cm}
  \includegraphics[width=.49\textwidth]{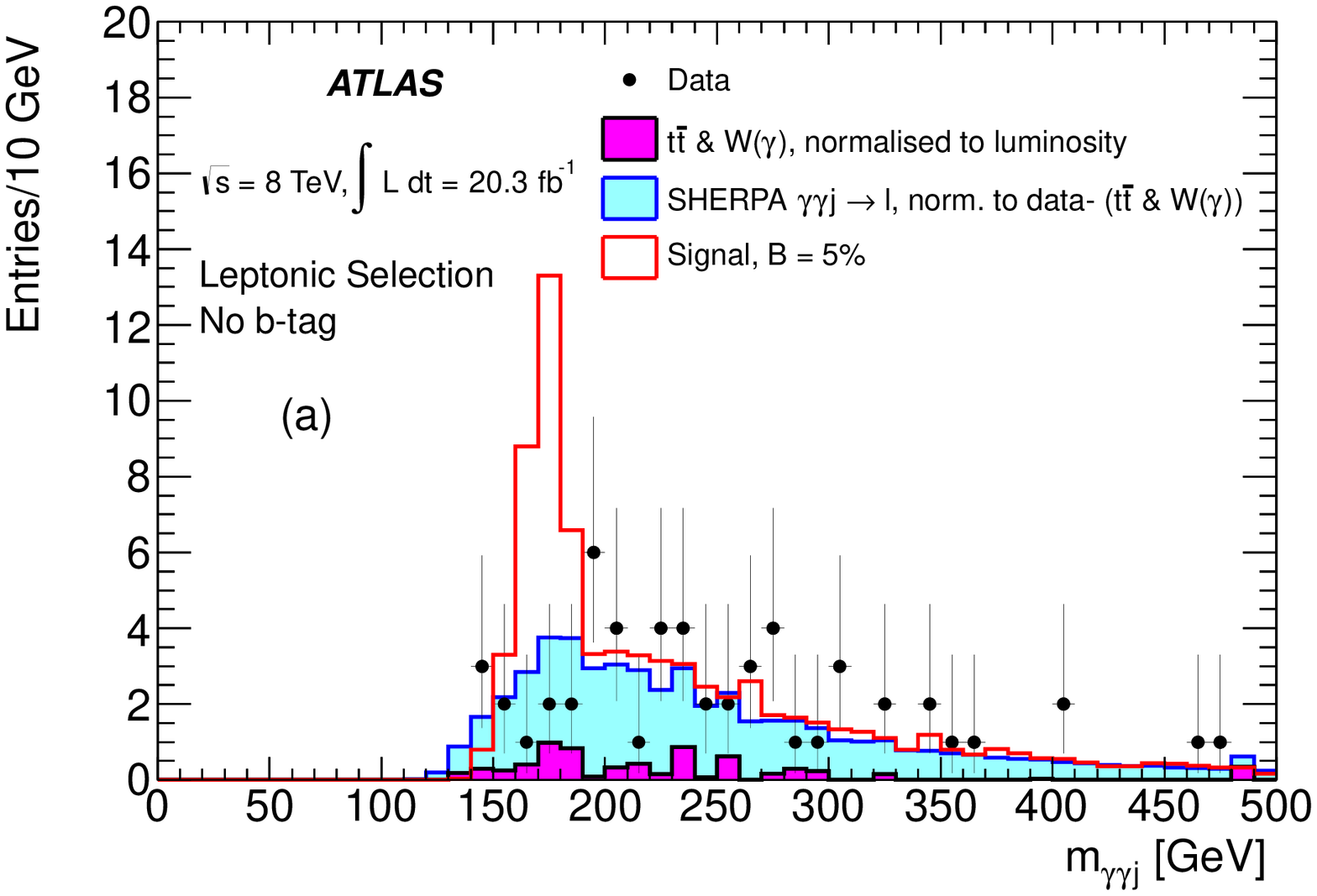}
\hspace{-0.4cm}
  \includegraphics[width=.49\textwidth]{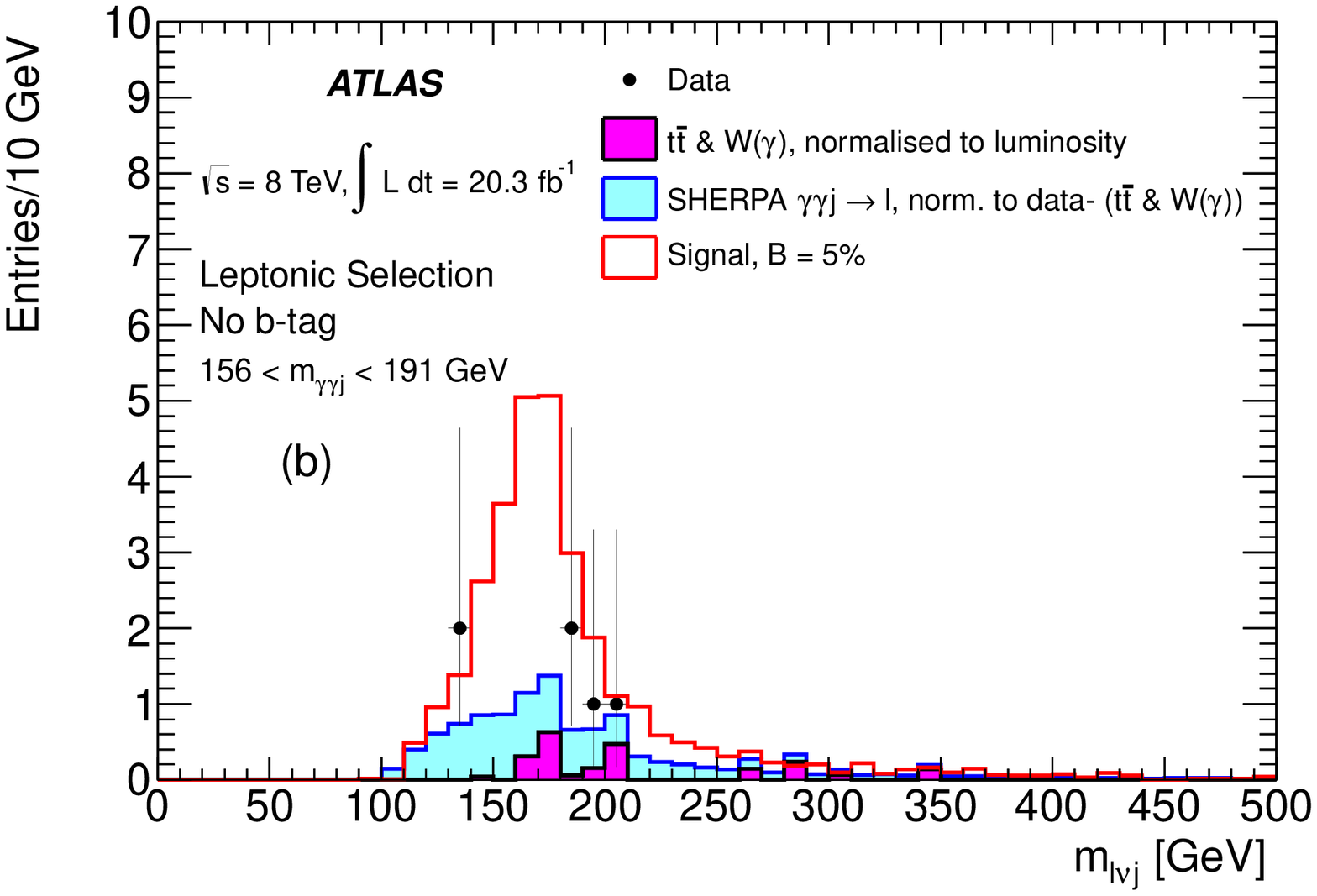}

\end{center}
\vspace{-0.5cm}
  \caption{(a) Distribution of the invariant mass $m_{\gamma\gamma j}$ ({\it Top1}) candidates , for events selected by the leptonic analysis (see text for details) in the $\sqrt{s}=8$ \tev\ data set, before $b$-tagging requirement. For each event there are two combinations, and both are displayed. (b) Distribution of the invariant mass $m_{l\nu j}$ for the complementary top candidates ({\it Top2}) decaying into one jet, a lepton and a neutrino; only combinations for 
which $m_{\gamma\gamma j}$ is between 156 and 191 \gev\ enter in the distribution.}
  \label{leptonic:fig:Mt12BT}
\end{figure}

Fig.~\ref{leptonic:fig:Mt12BT}(a) shows the invariant mass distribution of the {\it Top1} combinations (two entries per event)
for all selected events before the mass selections, and without the $b$-tagging requirement. 
In the signal sample, normalised to the expectation for a 5\% $t\rightarrow cH$ 
branching ratio, the narrow peak associated with the top quark is clearly visible,
as well as a tail at higher masses corresponding to the wrong combination of final state objects. The S$_{\gamma\gamma j\rightarrow \ell}$ sample, together with  the \nttb\ and $W(\gamma)$  contributions give a reasonable description of the data. The interval $\Delta m_{1}$ chosen for the $m_{1}$ selection is [156,191] \gev, as in the hadronic case. Only combinations for which $m_{1}$ fulfills the $\Delta m_{1}$ selection enter in the {\it Top2} distribution (Fig.~\ref{leptonic:fig:Mt12BT}(b)). Based on the width of the peak in the signal simulation, the interval $\Delta m_{2}$ is chosen to be [135,205] \gev, a little narrower than for the hadronic mode.

Table~\ref{tab:leptonic:stat} shows an overview of the leptonic selection at various stages of the analysis.
Inclusive \nttb\ and $W(\gamma)$ production, normalised to the luminosity of the data, are expected to contribute about 0.7 and 0.3 
events, respectively, and the S$_{\gamma\gamma j\rightarrow \ell}$ background about half an event. In the data a single event remains with a $\gamma\gamma$ mass of 147 \gev\ and a muon 
with $p_\mathrm{T}$ of 47 \gev. 
\begin{table}[h!]
\begin{center}
\caption{Efficiency (in percent) for $t\rightarrow cH$ signal simulation and numbers of events selected for data or expected (\nttb$\& W(\gamma)$, S$_{\gamma\gamma j\rightarrow \ell}$)  at different stages of the analysis, in the leptonic selection. The column denoted by ``\nttb$\& W(\gamma)$'' is normalised to the luminosity of the data. The column denoted by S$_{\gamma\gamma j\rightarrow \ell}$ is normalised to data after subtraction of the expected background from \nttb\ and $W(\gamma)$ at the ``2 photons + 1 lepton'' selection step. 
The uncertainties are statistical only for $t\rightarrow cH$ and \nttb$\& W(\gamma)$, but include the normalisation uncertainty for $S_{\gamma\gamma j\rightarrow \ell}$.}
\label{tab:leptonic:stat}
\vspace*{1em}
\footnotesize
\begin{tabular}{|l|c|cc|c|}
  \hline
                     & $t\rightarrow cH$ & \nttb\ $\& W(\gamma)$ & S$_{\gamma\gamma j\rightarrow \ell}$ & Data    \\ \hline
                                                 &  (\%)          & \multicolumn{3}{c|}{Events} \\ \hline \hline
$\gamma\gamma$ selection                         & 34.9           & 313.7       &             & 118500  \\
1 lepton                                         &  6.0           &  21.8       & 188.2       &    210  \\
N$_\mathrm{jets}\geq $ 2, $m_\mathrm{T} >$ 30 \gev &  3.8           &   3.4       &  18.8       &     30  \\
Mass requirements                                        &  1.9           &   1.2       &   3.5       &      4  \\ 
At least 1 $b$-tag                               &  1.3$\pm0.1$ & 0.9$\pm$0.5 & 0.5$\pm$0.2 &      1  \\\hline
\end{tabular}
\end{center}
\end{table}

\normalsize
The satisfactory agreement observed between data and background expectations indicates a good understanding of the 
background composition. 
However, as a sideband technique is used, only the shape of the background is relevant. 
Out of the three main contributions to the background (\nttb, $W(\gamma)$ and hadronic) the first two suffer from low statistics in 
the simulation. At earlier stages of the analysis, where more events are available, 
the distributions are smooth and exhibit a decreasing slope, compatible with the function used 
in Sec.~\ref{sec:shapes} to describe the background shape.

\section{Statistical analysis and results}
\label{sec:fit}
\label{sec:likelihood}
The parameter of interest is the branching ratio ${\cal B}$ of the decay $t \rightarrow c(u)H$.
A fit to the data  is performed using a likelihood function defined as the product of the likelihoods for the individual search 
channels, whose sensitivities as a function of ${\cal B}$ are given in Sec.~\ref{sec:channelsens}.
Hypothesised values of ${\cal B}$ are evaluated with a test statistic based on the profile likelihood ratio~\cite{asym}. 
In the hadronic selection, which combines the 7 and 8 \tev\ data, $m_{\gamma\gamma}$ is used as discriminating variable in the fit. The analysis in the leptonic selection is  
based on event counting in two $m_{\gamma\gamma}$ regions: the signal region (SR) from 122 to 129 \gev, 
and the control region (CR) from 100 to 122 \gev\ and from 129 to 160 \gev.
\par The theoretical uncertainties enter mainly through the \nttb\ production cross-section, the Higgs boson branching ratio to $\gamma\gamma$, the background 
due to SM Higgs production (Sec.~\ref{sec:othersources}) and the signal generator uncertainties (Sec.~\ref{sec:Expsys}). The experimental systematic uncertainties are detailed in Sec.~\ref{sec:Expsys}. All these uncertainties are introduced as nuisance parameters in the likelihood. 

\subsection{Expected signal event yields} 
\label{sec:channelsens}
The expected signal event yields in the three channels (hadronic 7 \tev, hadronic 8 \tev\ and leptonic 8 \tev) are estimated using the signal efficiencies given in Tables~\ref{tab:hadronic:stat} and~\ref{tab:leptonic:stat}, the \nttb\ production cross-sections at 7 and 8 \tev~\cite{TopCrossSection:2013,Botje:2011sn}, and the integrated luminosities of the corresponding data sets. They are listed in Table~\ref{tab:AnaCut:sensitivity}, where they are expressed in terms of the number of events expected for a $t \rightarrow cH$ branching ratio of 1\%. The same study with $t \rightarrow uH$ shows that the efficiency of the hadronic analysis is 1\% higher than for $t\rightarrow cH$, while  it is 6\% lower for the leptonic analysis. These variations are small enough to justify taking the same sensitivity for both modes.
\par The sensitivities are evaluated for a Higgs boson mass of 125.5 \gev~\cite{Aad:2013wqa}. A correction of -1\% (+1.5\%) is applied on the hadronic (leptonic) efficiency, obtained from a linear interpolation of the acceptances  estimated at particle level for simulations with masses of 125 and 126.8 \gev. Changing the top quark mass from 172.5 to 173.1 \gev\ increases the acceptances by about 1.6\% while the \nttb\ production cross-section is decreased by about 1.8\%. The net effect is thus neglected.  The effect of small differences between data and simulation in $b$-tagging and photon isolation efficiencies is also included in Table~\ref{tab:AnaCut:sensitivity}.
The generator and the experimental systematic uncertainties, not included in Table~\ref{tab:AnaCut:sensitivity}, are described in Sec.~\ref{sec:Expsys}. As pointed out above, the 7 \tev\ and the 8 \tev\ hadronic channels are treated as a single channel, whose combined expected event yield is $10.9\pm0.8$ events for ${\cal B} = 1\%$.

\begin{table}[h!]
\begin{center}
\caption{Expected signal efficiencies and event yields for a $t\rightarrow cH$ branching ratio of 1\% and $m_t$=172.5 \gev\ 
in the three analysis channels. The values used for the \nttb\ cross-section and the $H\rightarrow\gamma\gamma$ 
branching fraction are quoted for completeness.}
\label{tab:AnaCut:sensitivity}
\footnotesize
\vspace*{1em}
\begin{tabular}{|l|ccc|}
  \hline
 Selection                  &  \multicolumn{2}{c|}{Hadronic} & Leptonic \\
\hline
 Centre of mass energy                   & \multicolumn{1}{c|}{7 \tev\ }         &         \multicolumn{2}{c|}{ 8 \tev\ }    \\ \hline \hline
\nttb\ cross-section (pb)                  & \multicolumn{1}{c|}{177 $^{+10}_{-11}$} & \multicolumn{2}{c|}{253$^{+13}_{-15}$}\\
\hline
$H\rightarrow \gamma\gamma$ {\it Br} ($\%$) & \multicolumn{3}{c|}{0.23$\pm0.01$} \\
\hline
Signal efficiency (\%)                    & 4.2$\pm0.1$      & 4.0$\pm0.1$          & 1.3$\pm0.1$  \\ \hline
Exp. events for $\cal{B}$ = 1$\%$  & 1.6$\pm0.1$    & 9.3$\pm0.7$      & 3.0$\pm0.3$ \\ \hline
\end{tabular}
\end{center}
\end{table}

\subsection{Signal and background modelling}
\label{sec:shapes}

The shape of the signal diphoton mass distribution is similar to the shape used in the $H\rightarrow\gamma\gamma$ inclusive analysis~\cite{Aad:2013wqa}, for a signal mass hypothesis $m_H = 125.5 \gev$. It is described by the sum of a wide Gaussian and a Crystal Ball function with width $\sigma \simeq 1.7 \gev$, and differs slightly between the 7 and 8 \tev\ analyses. 
The fraction of the signal that falls into the SR is estimated to be $ \sim 90\%$. The same shape is used for the resonant background from SM Higgs boson production.

\subsubsection*{Background estimate for the hadronic channel} 
Due to low statistics the data distribution in the CR alone cannot be used to constrain the background shape.  
Instead, the diphoton mass spectrum from the S$_{\gamma\gamma j}$ sample (see Fig.~\ref{hadronic:fig:Mgaga}) smoothed using 
the algorithm of Ref.~\cite{Cranmer:2000du} is employed. 
Pseudo-data have been generated following this distribution, with on average 45.2 events
(given by the sum of the 38 data events in the CR and the associated SR contribution of 7.2 events assuming 
that the true probability density function is the smoothed one).
The corresponding \mgg\ spectra have been fitted with different parametrisations for the background shape. 
For fits with only the background,
a bias has been defined as the difference, in the SR, between  the true number of events and the number of events predicted by the fit. For 
fits including the signal, the bias is defined as the number of fitted signal
events. The criterion used to select a background parametrisation as valid is that these biases should be smaller than
10\% of the number of signal events at the expected limit ($\sim 6$ events). 
The background-only and the signal+background fits give consistent results. 
Both 2$^{nd}$ and 3$^{rd}$ order polynomial distributions satisfy the criterion, and the 2$^{nd}$ order polynomial was 
chosen. The associated bias with respect to the smoothed S$_{\gamma\gamma j}$ distribution is $\sim 0.6$
event. It is added as a systematic uncertainty in the final fit.

\subsubsection*{Background estimate for the leptonic analysis}
The background in the leptonic channel is estimated via a transfer factor $\alpha$, defined as the ratio of the background shape integral over the SR and its integral over the CR. The central value of $\alpha$ in the fit is given by the smoothed function used for the hadronic analysis,
$\alpha = 0.15$. For a flat \mgg\ distribution $\alpha \simeq$ 0.13 would be obtained.
A Gaussian constraint on $\alpha$ with a conservative width of 30\% is included in the likelihood function.

\subsection{Background from SM Higgs production}
\label{sec:othersources}
The estimate of the expected number of background events from SM Higgs production is obtained by combining the cross-sections for Higgs boson production via the $ggF$, VBF, $WH$, $ZH$ and \nttb$H$
processes~\cite{LHCHiggsCrossSectionWorkingGroup:2011ti, LHCHiggsCrossSectionWorkingGroup:2012vm}, assuming they all follow the SM predictions, the integrated luminosities of the 7 \tev\ and 8 \tev\ data sets, and the event selection efficiencies determined using full simulation for each production mode (see Sec.~\ref{sec:MC}). The uncertainties on the cross-sections are obtained by a linear sum of the renormalisation and factorisation scale uncertainties on one hand, and of the parton distribution functions and $\alpha_s$ uncertainties on the other hand, as they appear in Refs.~\cite{LHCHiggsCrossSectionWorkingGroup:2011ti, LHCHiggsCrossSectionWorkingGroup:2012vm}.  The VBF process gives a negligible contribution.
In the absence of fully simulated samples for the $tH$ production, for which the cross-section was only recently 
calculated~\cite{TopHiggs:Biswas, TopHiggs:Grojean},
the acceptance is obtained from particle level simulation, scaled by the ratio of acceptances for full and particle level simulations obtained for topologically similar final states ($tcH$ and \nttb$H$ final states were used).
\par Since the $ggF$ and $WH$ processes produce a Higgs boson with a small number of jets, among which there is 
in general no $b$-jet, an additional systematic uncertainty is added. For the $ggF$ mode, several variations of POWHEG+PYTHIA8~\cite{MINLO} with up to three partons at the matrix element level were compared, from which a 40\% uncertainty was deduced. For the $WH$ process, which is simulated at LO, the uncertainty is enlarged to 100\%.  The uncertainty on $ZH$ is not increased as this process produces $b$-jets from the $Z$-boson decay. 
\par In total, for the hadronic selection, the background from SM Higgs production is 0.24$\pm$0.05 event at 8 \tev\ and 0.04$\pm$0.01 at 7 \tev, with uncertainties taken as fully correlated. The largest contribution is from the \nttb$H$ mode, which represents about 60\% of the total. In the leptonic selection, the total background due to SM Higgs production amounts to 0.05$\pm$0.01 event, and 90\% of this background arises from  \nttb$H$ production.

\subsection{Experimental systematic uncertainties}
\label{sec:Expsys}
\par The experimental systematic uncertainties are listed in Table~\ref{tab:Sumuncer}.
\begin{table}[h!]
\begin{center}
\caption{Summary of experimental and generator (see text) uncertainties on the signal and SM Higgs boson background yields (in percent, per event).
The last row gives the sum in quadrature of all these uncertainties.}
\label{tab:Sumuncer}
\footnotesize
\vspace*{1em}
\begin{tabular}{|l|cc|c|}
  \hline
  Selection            &  \multicolumn{2}{c|}{Hadronic} & Leptonic \\
\hline
  Centre of mass energy   &   7 \tev\               &       8 \tev\               & 8 \tev\ \\
  \hline \hline
  Trigger efficiency     & $\pm$0.2& $\pm$0.5& $\pm$0.5                    \\
  Photon identification                    & $\pm$9.3    & $\pm$4.6      & $\pm$2.4     \\
  Photon isolation       & \multicolumn{2}{c|}{$\pm$3.0}     & $\pm$1.0       \\[0.8ex]
  Jet Energy Scale                    & $\pm$5.4  & $^{+\mbox{\footnotesize 7.4}}_{-\mbox{\footnotesize 4.5}}$& $^{+\mbox{\footnotesize 3.2}}_{-\mbox{\footnotesize 2.8}}$  \\[0.8ex]
  Jet Energy Resolution                    & \multicolumn{2}{c|}{$\pm$0.2}&$\pm$0.2                   \\
  Jet Vertex Fraction    & \multicolumn{2}{c|}{$\pm$1.0} &    $\pm$1.0               \\
  $b$-tagging            & $\pm$3.5     & $\pm$4.8        & $\pm$5.2    \\
  Lepton reco./ID/scale  & --             & --                & $\pm$0.6         \\[0.8ex]
 \met\ scale            & --             & --                &$^{+\mbox{\footnotesize 1.4}}_{-\mbox{\footnotesize 0.4}}$ \\[0.8ex]
\hline
\rule[-0.0cm]{0cm}{.4cm}
 ISR/FSR      & \multicolumn{2}{c|}{$^{+\mbox{\footnotesize 7.0}}_{-\mbox{\footnotesize 3.0}}$ }    & $^{+\mbox{\footnotesize 8.0}}_{-\mbox{\footnotesize 2.0}}$     \\[0.8ex]  
Underlying event       & \multicolumn{2}{c|}{$\pm$3.5}     & $\pm$1.8   \\
  \hline \hline
\rule[-0.0cm]{0cm}{.4cm}
Combined uncertainty &   $^{+\mbox{\footnotesize 14.1}}_{-\mbox{\footnotesize 12.6}}$ & $^{+\mbox{\footnotesize 13.1}}_{-\mbox{\footnotesize 9.8}}$& $^{+\mbox{\footnotesize 10.6}}_{-\mbox{\footnotesize 7.1}}$  \\[0.8ex]
\hline
\end{tabular}
\end{center}
\end{table}
\normalsize
\begin{itemize}
\item The uncertainties related to photons are described in Sec.~\ref{sec:sample} for the trigger efficiency, and in Sec.~\ref{sec:photonID} for the photon identification and isolation.
\item The systematic uncertainty associated with the Jet Energy Scale (JES) is determined by changing by one standard deviation, in each direction and one at a time, each of the parameters to which the energy scale is sensitive. 
The most sensitive parameters are associated with pile-up and  jet flavour. At 8 \tev\ the quadratic sum of the uncertainties obtained from all variations gives a total effect on the expected signal yield 
of (+7.4\%,$\mathrm{-}4.5$\%). At 7 \tev\ the global effect is more symmetric ($\pm 5.4\%$). The smaller JES uncertainty in the leptonic channel was obtained in the same way and includes its impact on \met. The same methodology was used for the Jet Energy Resolution (JER), whose uncertainty has a smaller impact on the signal yield.
\item The systematic uncertainty associated with the JVF selection is estimated by varying the corresponding requirement within the boundaries resulting from 
dedicated studies~\cite{JVFsys}. It amounts to 1\% for both the hadronic and the leptonic selections. The same uncertainty is also used at 7 \tev.
\item In order to take into account the small differences in $b$-tagging efficiency between data and simulation for each
 jet flavour (light, charm and bottom-quark jets)~\cite{ATLAS-Btagging}, the nominal values of the associated scale factors 
are included in the event weights of the simulated samples. 
Replacing the nominal scale factors by the values obtained when adding (subtracting) their uncertainty induces variations of the expected 
signal yield of the order of 5\%.
\item The uncertainty associated with the lepton energy scale, identification and reconstruction efficiency, averaged for electrons and muons, is 0.6\%. 
\item The uncertainty, of about 1\%, associated with \met\ was obtained with the same methodology as that used for the jet energy scale, 
applied to low-$E_\mathrm{T}$ topological clusters included in the estimate of \met\ and which are not associated with any 
of the objects used to reconstruct the final state.
\end{itemize}
The generator uncertainties are evaluated as follows:
\begin{itemize}
\item The uncertainty labelled ``ISR/FSR'' in Table~\ref{tab:Sumuncer} corresponds to the variation of the signal acceptance 
observed at particle level when the parameters governing QCD initial and final state radiation in PYTHIA6 are varied within the allowed 
range~\cite{TunesVariation}.  
\item The systematic uncertainty associated with the underlying event modelling is estimated by scaling, in the simulation,
the transverse momenta of particles produced at $|\eta|>2$ within the range allowed by the differences between 
tunes~\footnote{The particle flow observed in various data samples for $|\eta| < 2$ is well described by standard QCD PS and MI tunes.} 
and re-estimating the selection efficiency. 
\end{itemize}

\subsection{Results}
\par A fit using the likelihood described at the beginning of this section is performed
on the selected data sample, consisting of 50 events in the hadronic channel 
and one event in the leptonic channel.
\begin{figure}[h!]
\begin{center}
\hspace{-0.6cm}
\includegraphics[width=.70\textwidth]{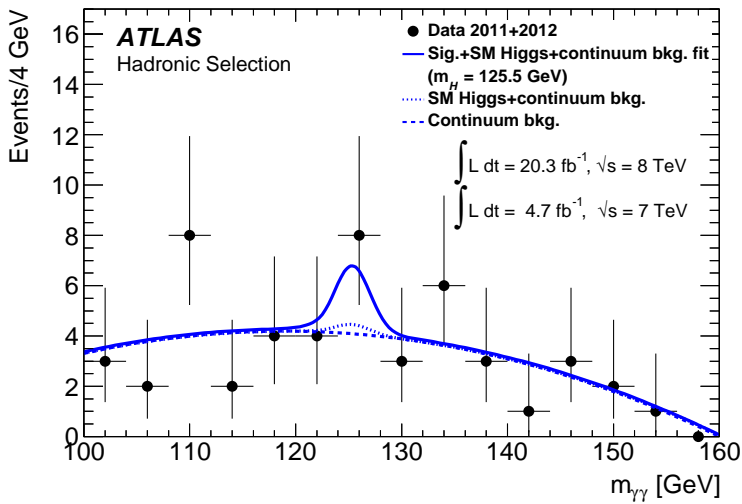}
\hspace{-0.4cm}
\end{center}
\vspace{-0.5cm}
\caption {Distribution of \mgg\ for the selected events in the hadronic channel. The result of a fit to the data of the sum of a signal 
component with the mass of the Higgs boson fixed to $m_H = 125.5 \GeV$ and a background component (dashed) described by a second-order 
polynomial is superimposed. The small contribution from SM Higgs boson production, included in the fit, is also shown (difference between 
the dotted and dashed lines).}
\label{fig:result}
\end{figure} 
 \par The diphoton mass spectrum in the hadronic channel is shown in 
Fig.~\ref{fig:result}, together with the fitted background shape and the signal shape for a Higgs boson mass fixed at 125.5 \gev.
The fitted branching ratio is ${\cal B} = 0.22^{+0.31}_{-0.26}\%$, which corresponds to a total number of signal events (hadronic and leptonic) of $3.1^{+4.3}_{-3.7}$. The probability that the background can produce a fluctuation greater than or equal to the excess observed in data is 18\%.
As no significant signal is found, limits on the $t \rightarrow cH$ and $t \rightarrow uH$ branching ratios are set based on the CL$_\mathrm{s}$ prescription~\cite{Read:2002hq}.

\par
\begin{figure}[h!]
\begin{center}
\includegraphics[width=.75\textwidth]{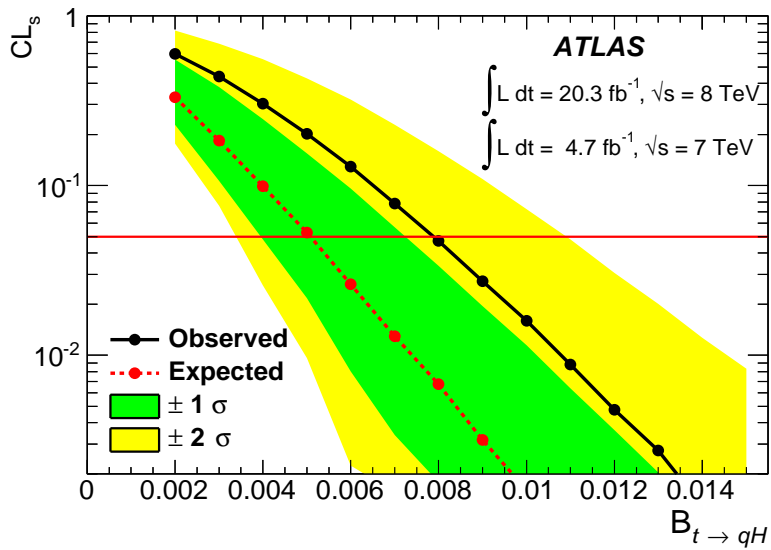}
\end{center}
\vspace{-0.8cm}
\caption{Evolution of CL$_\mathrm{s}$ as a function of the branching fraction $\cal{B}$ of the $t \to qH$ decay 
for the observation of a signal at 125.5 \gev\ (solid line) and the expectation in the absence of signal 
(dashed line). The 1 and 2 $\sigma$ uncertainty bands around the expected curve are also shown.}
\label{fig:cls}
\end{figure}

 The evolution of the signal confidence level CL$_\mathrm{s}$ as a function of the branching fraction ${\cal B}$ for $t\rightarrow qH$ 
is shown in Fig.~\ref{fig:cls}. 
Pseudo-experiments have been used to determine
the distributions of the test statistic under the signal+background and the background-only hypotheses. 
The green and yellow areas represent the one and two standard deviation bands around the expectation. The observed (expected) 
limit on $\cal{B}$ is 0.79 (0.51)\% at the 95\% confidence level. The observed limit is not as stringent as the expectation 
due to a slight excess over the total background expectation in the vicinity 
of \mgg $\sim$ 126 \gev, as seen in Fig.~\ref{fig:result}. From this limit, an upper limit on the $\lambda_{tcH}$ coupling of 0.17 
was obtained, with an expected value of 0.14. As the analysis is equally sensitive to the $t\rightarrow uH$ and $t\rightarrow cH$ modes, the limit obtained on the couplings can be written as $\sqrt{\lambda_{tcH}^{2} + \lambda_{tuH}^{2}} < 0.17$, with an expectation of 0.14.

\section{Conclusions}
\par  The FCNC $t \rightarrow qH$ decay, followed by $H\rightarrow\gamma\gamma$, has been searched for in a 
data set  of proton-proton  collisions recorded by the ATLAS experiment, consisting of 4.7 fb$^{-1}$ at $\sqrt{s}=7$ \tev\ and 20.3 fb$^{-1}$ at $\sqrt{s}=8$ \tev.
\par Candidate events were selected by requiring the presence of two high-$E_\mathrm{T}$ 
isolated photons. Further selection criteria included the presence of 
four jets (at least one $b$-tagged) for the hadronic selection, or two jets (at least one $b$-tagged), 
\met\ and an isolated lepton for the leptonic selection, plus kinematic conditions designed to enhance the fraction of events with a \nttb\ topology.

\par A sideband technique was used to constrain the background, and an expected upper limit on the $t\rightarrow cH$ 
decay branching ratio in the absence of signal of 0.51\% was calculated. No statistically significant excess was observed 
in the data, and a limit of 0.79\% was set at the 95\% 
confidence level for $m_H = 125.5 \gev$. From this limit, an upper limit on the $\lambda_{tcH}$ coupling of 0.17 
was obtained, with an expected value of 0.14. As the analysis is equally sensitive to the $t\rightarrow uH$ 
and $t\rightarrow cH$ modes, the limit obtained can more generally be expressed as $\sqrt{\lambda_{tcH}^{2} + \lambda_{tuH}^{2}} < 0.17$.

\section*{Acknowledgments}


We thank CERN for the very successful operation of the LHC, as well as the
support staff from our institutions without whom ATLAS could not be
operated efficiently.

We acknowledge the support of ANPCyT, Argentina; YerPhI, Armenia; ARC,
Australia; BMWF and FWF, Austria; ANAS, Azerbaijan; SSTC, Belarus; CNPq and FAPESP,
Brazil; NSERC, NRC and CFI, Canada; CERN; CONICYT, Chile; CAS, MOST and NSFC,
China; COLCIENCIAS, Colombia; MSMT CR, MPO CR and VSC CR, Czech Republic;
DNRF, DNSRC and Lundbeck Foundation, Denmark; EPLANET, ERC and NSRF, European Union;
IN2P3-CNRS, CEA-DSM/IRFU, France; GNSF, Georgia; BMBF, DFG, HGF, MPG and AvH
Foundation, Germany; GSRT and NSRF, Greece; ISF, MINERVA, GIF, I-CORE and Benoziyo Center,
Israel; INFN, Italy; MEXT and JSPS, Japan; CNRST, Morocco; FOM and NWO,
Netherlands; BRF and RCN, Norway; MNiSW and NCN, Poland; GRICES and FCT, Portugal; MNE/IFA, Romania; MES of Russia and ROSATOM, Russian Federation; JINR; MSTD,
Serbia; MSSR, Slovakia; ARRS and MIZ\v{S}, Slovenia; DST/NRF, South Africa;
MINECO, Spain; SRC and Wallenberg Foundation, Sweden; SER, SNSF and Cantons of
Bern and Geneva, Switzerland; NSC, Taiwan; TAEK, Turkey; STFC, the Royal
Society and Leverhulme Trust, United Kingdom; DOE and NSF, United States of
America.

The crucial computing support from all WLCG partners is acknowledged
gratefully, in particular from CERN and the ATLAS Tier-1 facilities at
TRIUMF (Canada), NDGF (Denmark, Norway, Sweden), CC-IN2P3 (France),
KIT/GridKA (Germany), INFN-CNAF (Italy), NL-T1 (Netherlands), PIC (Spain),
ASGC (Taiwan), RAL (UK) and BNL (USA) and in the Tier-2 facilities
worldwide.

\bibliographystyle{JHEP}
\bibliography{noteJHEP}

\clearpage
\begin{flushleft}
{\Large The ATLAS Collaboration}

\bigskip

G.~Aad$^{\rm 84}$,
B.~Abbott$^{\rm 112}$,
J.~Abdallah$^{\rm 152}$,
S.~Abdel~Khalek$^{\rm 116}$,
O.~Abdinov$^{\rm 11}$,
R.~Aben$^{\rm 106}$,
B.~Abi$^{\rm 113}$,
M.~Abolins$^{\rm 89}$,
O.S.~AbouZeid$^{\rm 159}$,
H.~Abramowicz$^{\rm 154}$
% ...
\bigskip
\\
% ...
$^{11}$ Institute of Physics, Azerbaijan Academy of Sciences, Baku, Azerbaijan\\
% ...
$^{84}$ CPPM, Aix-Marseille Universit{\'e} and CNRS/IN2P3, Marseille, France\\
% ...
$^{89}$ Department of Physics and Astronomy, Michigan State University, East Lansing MI, United States of America\\
% ...
$^{106}$ Nikhef National Institute for Subatomic Physics and University of Amsterdam, Amsterdam, Netherlands\\
% ...
$^{112}$ Homer L. Dodge Department of Physics and Astronomy, University of Oklahoma, Norman OK, United States of America\\
$^{113}$ Department of Physics, Oklahoma State University, Stillwater OK, United States of America\\
% ...
$^{116}$ LAL, Universit{\'e} Paris-Sud and CNRS/IN2P3, Orsay, France\\
% ...
$^{152}$ Institute of Physics, Academia Sinica, Taipei, Taiwan\\
% ...
$^{154}$ Raymond and Beverly Sackler School of Physics and Astronomy, Tel Aviv University, Tel Aviv, Israel\\
% ...
$^{159}$ Department of Physics, University of Toronto, Toronto ON, Canada\\
% ...
\end{flushleft}


\end{document}